\documentclass{emulateapj}

\usepackage{amssymb,amsmath} 

\newcommand\beq{\begin{equation}}
\newcommand\eeq{\end{equation}}
\newcommand\beqar{\begin{eqnarray}}
\newcommand\eeqar{\end{eqnarray}}

\newcommand{\cref}{C_{\rm ref}}

\begin{document}

\title{A New Recipe for Obtaining Central Volume Densities \\  of  Prestellar Cores from Size Measurements} 

\author{Konstantinos Tassis\altaffilmark{1}, 
Harold W. Yorke\altaffilmark{1}
}

\altaffiltext{1}{Jet Propulsion Laboratory, California Institute of Technology, Pasadena, CA 91109, USA}

\begin{abstract}

We propose a simple analytical method for estimating the central volume density of prestellar molecular cloud cores from their column density profiles. Prestellar cores feature a flat central part of the column density and volume density profiles {\em of the same size} indicating the existence of a uniform density inner region. The size of this region is set by the thermal pressure force which depends only on the central volume density and temperature of the core, and can provide a direct measurement of the central volume density. Thus a simple length measurement can immediately yield a central density estimate independent of any dynamical model for the core and without the need for fitting. Using the radius at which the column density is 90\% of the central value as an estimate of the size of the flat inner part of the column density profile yields an estimate of the central volume density within a factor of 2 for well resolved cores. 
\end{abstract}


\section{Introduction}\label{intro}

Prestellar molecular cloud cores are the very first stages of the star formation process, and as such they are the subject of intense observational investigations, as their properties can reveal important clues about the initial conditions of star formation. Column density profiles of prestellar cores encode information about their structure, stability, and evolutionary stage. They are measured using various methods, including mm continuum emission (e.g., Ward-Thompson et al.~1999), mid-IR absorption (e.g., Bacmann et al.~2000), dust extinction and reddening of background stars in near-IR (e.g., Alves et al.~2001;  Kandori et al.~2005), and flux measurements in optically thin lines (e.g., Tafalla et al.~2002). These profiles typically feature a uniform density  inner region (Ward-Thompson et al.~1994; Ward-Thompson et al.~1999; Bacmann et al.~2000; Shirley et al.~2000; Schnee et al.~2010), followed by a power law decline. 

Detailed fits of various equilibrium (e.g., Bonnor-Ebert spheres, Bonnor 1956; Ebert 1955) or dynamically evolving models (e.g., Crutcher et al.~1994; Ciolek \& Basu 2000) to column density profiles can provide estimates for a variety of core properties. However these detailed fits are often computationally expensive to produce, as profiles for various parameter values are typically only available as numerical solutions. Although analytic profile fits can alleviate this problem (Dapp \& Basu 2009), the results of full profile fits are necessarily model and geometry dependent. 

Here we propose a simple and model-insensitive way to estimate the {\em central value of the volume density} of a starless molecular cloud core  from its column density profile.  Although this technique provides the value of only a single quantity, it has the advantage that it is computationally inexpensive, as it requires no fitting but only an estimate of the size of the uniform density inner region; it is insensitive to geometry or amount of magnetic support; and the quantity it estimates is a very important one, as the central {\em volume} density  sets the evolutionary stage of the core within the context of any specific dynamical model, and so it is necessary for further core modeling, including velocity, magnetic field, and  chemistry. 

The reason for the robustness of this technique is the simplicity of the physics behind the formation of the uniform density inner region: the flat inner region of the column density profile corresponds to the scale within which thermal pressure erases any inhomogeneities, the thermal lengthscale. It is therefore expected to be present independently of the details of the core dynamics, such as the amount of magnetic support of the core. Its extent is also not very sensitive to geometry effects or core orientation, as in the uniform density inner region pressure forces, even in magnetic cores, are much more prominent than further out, and the geometry is  closer to spherical than at larger scales. Since the thermal lengthscale depends only on the value of the central volume density of the core and the core temperature, which, in prestellar cores, typically varies by only a factor of $\sim 2$, a measurement of the thermal lengthscale, using the size of the flat inner region of the column density profile, can yield an estimate of the central volume density with comparable (factor of $\sim 2$) accuracy. With independent estimates of the kinetic temperature (e.g. from line intensity ratios), the accuracy on the central volume density can improve further.  

In \S \ref{method} we describe our proposed technique in detail, we demonstrate its robustness, and we discuss systematic uncertainties entering estimates of the central volume density obtained this way. In \S \ref{application} we apply our method and obtain estimates for the central volume density of the starless core B68 and the prestellar core L1544. We discuss our findings in \S \ref{discussion}. 

\section{Method}\label{method}

\begin{figure*}
\plotone{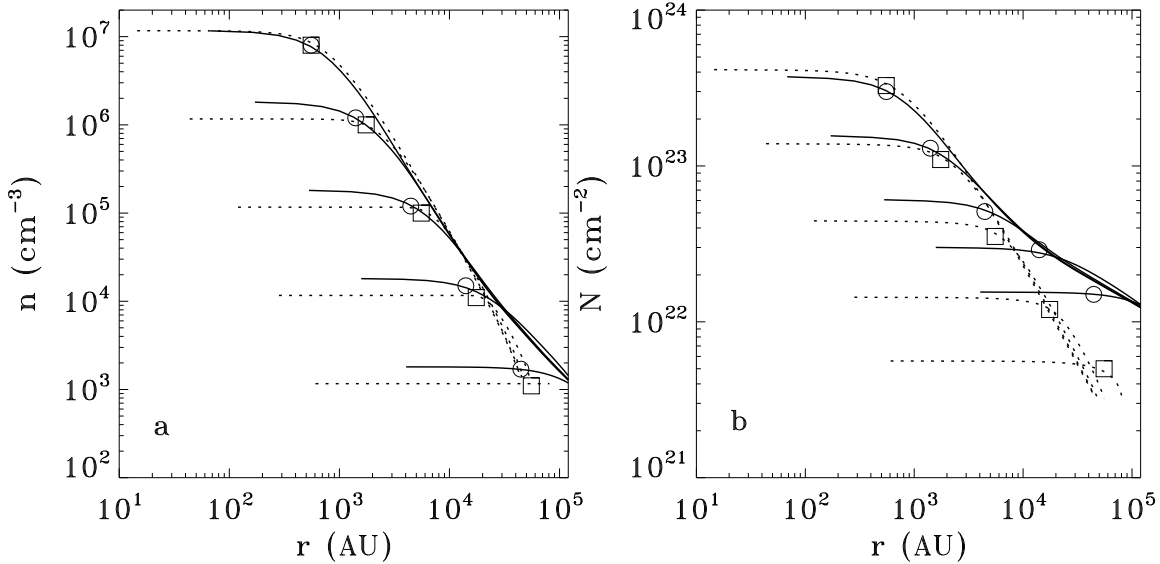}
\caption{\label{comp_dynamics_rad} 
Radial profiles of the number volume density $n$ (panel a); 
  and column density  (panel b). Dotted lines: non-magnetic model; solid lines: magnetic model. Each curve corresponds to a snapshot in time when the volume density is an order of magnitude higher than the previous one. Time elapsed per snapshot in magnetic model: 4.85 Myr, 5.61Myr, 5.74Myr, 5.77My; in nonmagnetic model: 0.89 Myr, 1.02Myr, 1.05Myr, 1.06Myr. The fifth (lowest-density) snapshot in both cases is the initial configuration of each model. The thermal lengthscale (Eq.~\ref{lc}) is indicated with a circle for the magnetic model and a square for the nonmagnetic model. }
\end{figure*}

We first give a simple demonstration of the inner flatness of the column density and volume density profiles, and the similarity of the observed extent of the uniform-density inner region in the volume density and column density profiles, for two dynamical models of prestellar molecular cloud cores. The first is a spherically symmetric non-magnetic dynamical model. The second is an axially symmetric magnetic model, where the initial mass-to-flux ratio is equal to the critical value for collapse (Mouschovias \& Spitzer 1976). Details for these simulations are given in Tassis et al. (2010, in preparation); the two models discussed here are the ``reference'' non-magnetic and magnetic models in that work, respectively. 

In Fig.~\ref{comp_dynamics_rad} we plot radial profiles of the volume density (left panel) and the column density (right panel),  for the non-magnetic (dotted lines) and magnetic (solid lines) models, for different time snapshots. Each snapshot corresponds to a time when the central volume density is an order of magnitude higher than the previous one. The time elapsed between the initialization of each simulation and each snapshot are given in the caption.

\begin{figure}
\plotone{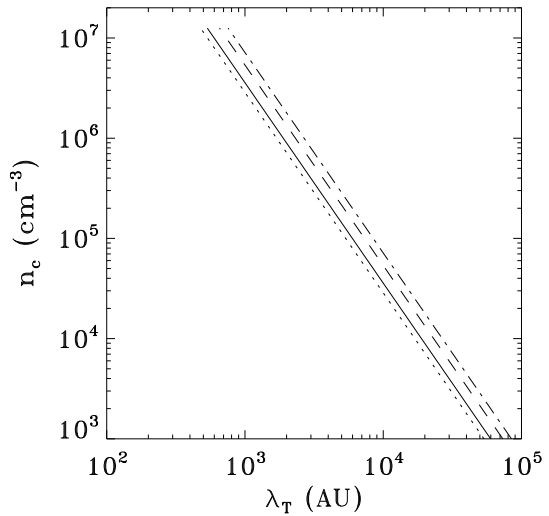}
\caption{\label{lT} Central density as a function of thermal lengthscale (Mouschovias 1991) for four different core temperatures. Solid line: $T=10$K; dotted line: $8$K; dashed line: $15$K; dot-dashed line: $20$K. Because observed core temperatures only span a small range, observing $\lambda_T$ can give us the central volume density. }
\end{figure}

The radial profiles of the volume density and the column density have similar properties, and feature a flat inner region, the boundary of which is indicated with a circle for the magnetic model and a square for the non-magnetic model, followed by a power-law decline with increasing radius. The angular size of the {\em column density} profile flat inner region is observationally measurable through a variety of techniques (see discussion in \S \ref{intro}).  Combined with an estimate for the distance of the core, such observations can yield the physical size of the column density profile flat inner region, which, however, is of the same size as the {\em volume density profile flat inner region}, as can be directly seen by comparing the left and the right panels of Fig.~\ref{comp_dynamics_rad}. 
The latter corresponds to the thermal lengthscale (Mouschovias 1991), which is equivalent to the critical Bonnor-Ebert sphere (Bonnor 1956; Ebert 1955), and it depends on the temperature and the central density of the core:
\begin{equation}\label{lc}
\lambda_T = 6\times 10^4 
\left(\frac{T}{10{\rm \, K}}\right)^{1/2}
\left(\frac{10^3 {\rm \, cm^{-3}}}{n_c}\right)^{1/2} {\rm \, AU}.
\end{equation}

Because the temperature does not vary much between prestellar cores, measuring $\lambda_T$ is a proxy to {\em measuring the central volume density} of a collapsing core. The very modest (factor of $\sim 2$) variation of $n_c$ for a given value of $\lambda_T$ due to the corresponding factor of $\sim 2$ spread in possible core temperatures is shown in Fig.~\ref{lT}. If the core temperature can also be independently measured, then the accuracy with which $n_c$ can be obtained through a measurement of $\lambda_T$ improves even further. Equation (\ref{lc}) allows one to obtain an actual {\em volume density} measurement directly from a {\em size} measurement. This result is independent of dynamical model and core geometry, as it 
is a very basic consequence of thermal pressure.

\begin{figure}
\plotone{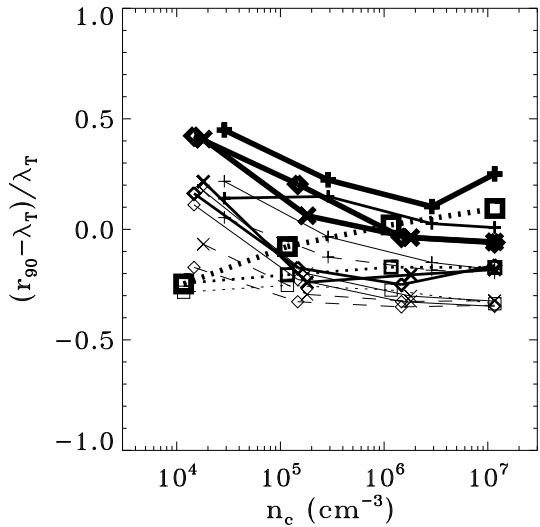}
\caption{\label{ERR} Fractional error, for various dynamical models and beam sizes, in the calculation of the size of the thermal lengthscale from the column density profile, when using the criterion $N(\lambda_T) = 0.9 N(0)$. Thin dotted line: non-magnetic spherical model; thin solid lines: magnetic models, viewed edge-on, symbols indicate the initial mass to magnetic flux ratio (diamond: subcritical, magnetically supported; x: critical; +: supercritical). Dashed curves are magnetic models viewed face-on (symbols as for thin solid curves). Normal solid and dotted lines correspond to a beam size half the angular size of $\lambda_T$, for magnetic and non-magnetic models respectively. Thick solid and dotted lines lines correspond to a beam size equal to $\lambda_T$, for magnetic and non-magnetic models respectively.}
\end{figure}

In practice, we recommend determining $\lambda_T$ from the column density profile $N(r)$ through the criterion 
\begin{equation}\label{thecrit}
N(\lambda_T) = 0.9 N(0)\,.
\end{equation}
This criterion is independent of the absolute calibration of the column density profile and is not affected by any uncertainties in this calibration. The fractional value of 0.9 at the ``edge'' of the uniform-density inner region is optimized so as to minimize sensitivity to  the details of the dynamical model, beam size, and core orientation with respect to the line of sight.
 
\begin{figure}
\plotone{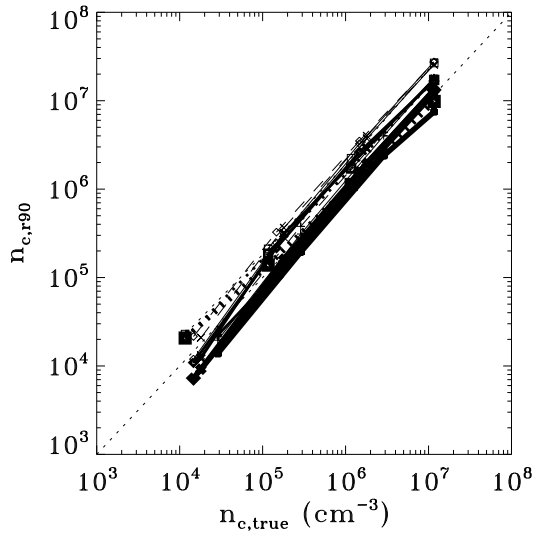}
\caption{\label{ERR2}Estimated vs true value of the central volume density. Line types and symbols as in Fig.~\ref{ERR}. The long dotted line denotes $n_{c,r90} = n_{c,\rm true}$.  }
\end{figure}

The robustness of this criterion,  as well as a quantitative estimate of the systematic uncertainties in the determination of $\lambda_T$ and $n_c$ entering through these considerations, are demonstrated in Figs.~\ref{ERR} and \ref{ERR2}. Figure \ref{ERR} shows the fractional error in the determination of  $\lambda_T$ when the criterion of Eq.~(\ref{thecrit}) is used, for various dynamical models and beam sizes, detailed in the caption. In magnetic cores the criterion of Eq.~(\ref{thecrit}) tends to overestimate $\lambda_T$ at lower central densities, while 
%as magnetic profiles are less steep outside the uniform column density region. This
the trend is reversed at higher central densities (and later times).
%as can be seen in the right panel of Fig.~\ref{comp_dynamics_rad}, the column density profile right outside the flat part of the profile in magnetic runs is significantly steeper than at very early snapshots. 
The physical reason behind this systematic trend is the extra magnetic support perpendicular to the field lines in magnetic cores, which is decreased at higher densities. The beam size does not significantly degrade the accuracy of the $\lambda_T$ determination, as long as the flat part of the profile is resolved. Even with a beam size comparable to the size of $\lambda_T$ (thick lines in Fig.~\ref{ERR}), the fractional error does not exceed $50\%$. The difference between face-on (dashed lines) and edge-on (solid lines) orientations of magnetic cores is small. 

The corresponding error introduced  to the estimate of $n_{c}$ through these systematic uncertainties affecting the determination of $\lambda_T$ is shown in Fig. \ref{ERR2}. If we do not consider marginally resolved inner core regions (i.e. beam sizes $\sim \lambda_T$; thick lines), the deviation from $n_{c,\rm true}$ is always smaller than a factor of 2. Even with the inclusion of the large beam size cases, the deviation does not exceed a factor of 3.

\section{Application}\label{application}

As an example of our proposed recipe, we apply the simple technique described above in two well-studied molecular cloud cores, B68 and L1544.

For L1544 we can read off of both the 450 and the 850 $\mu$ m  profiles of Shirley et al. (2000) through a simple linear interpolation between points that the column density (intensity) declines to 90\% its central value at a distance of $r_{90} \sim 1400$ AU from the center of the core. Using a temperature of 12 K (e.g., Williams et al. 1999), we obtain an estimate of the central volume density of $n_{c} \sim 2 \times 10^6$ cm$^{-3}$. This is significantly higher than the estimate of Williams et al.~ (1999), however it is consistent with the higher estimate of $\sim 10^6$ cm$^{-3}$ of  Ward-Thompson et al. (1999) and it is in excellent agreement with the value obtained by Ciolek \& Basu (2000) through detailed fitting of a magnetic model to this core.

For B68 we use the data of Alves et al. (2001) to obtain an estimate of $r_{90} \sim 2700$ AU. For $T\approx 10$K (Hatzel et al.~2002; Lai et al.~2003) we obtain $n_c \sim 5\times 10^5$ cm$^{-3}$. This value is consistent, within systematic uncertainties discussed above, with the value of $\sim 3\times10^5$ obtained by Dapp \& Basu (2009) by detailed fitting to their analytical approximation of a Bonnor-Ebert profile.

\section{Discussion}\label{discussion}

We have shown that, through a very simple analytic recipe that does not require fitting, it is possible to obtain an estimate of the central volume density of a starless molecular cloud core from its column density profile. 

For the optimal application of this technique, the beam size used should be small enough so that the flat part of the column density profile is well resolved, and the temperature of the core should be independently known. Under these conditions, through a simple measurement of the extent of the uniform column density region of the profile, we can estimate the value of the thermal lengthscale and from there the central {\em volume} density, a valuable input for further modeling, and an evolutionary ``clock'' for any specific dynamical model. This procedure returns a {\em model-insensitive} estimate for the central volume density within a factor of two. 

Uncertainties in the estimate of the central density in addition to variations in the shape of the profile due to the varying amount of magnetic support can arise through the following effects. 
\begin{itemize}
\item[(i)] If the temperature of the core under consideration is not independently known, additional systematic uncertainty, proportional to the uncertainty in the value of $T$, will enter our final estimate of $n_c$. 
\item[(ii)] Although the temperature gradient in the inner flat part of the profile is expected to be small (e.g., Tafalla et al.\ 2002), any deviations from isothermality will similarly induce uncertainties in the determination of the density profile and the estimate of $n_c$. 
\item[(iii)] Deviations of cores from spherical symmetry will impact the shape of the profile even close to the flat inner part and can affect the estimate of $n_c$. Cores typically do have aspect ratios of $\sim 2$, and although we have considered here flattened cores in the case of magnetic support, it is possible that aspherical cores may arise for other reasons; however the uncertainty induced on $n_c$ in such cases is not expected to exceed that which we have explicitly calculated for cores deviating from sphericity due to varying amounts of magnetic support, which also exhibit aspect ratios of 2 or higher. 
\item[(iv)] The calculation of a central volume density requires the knowledge of a physical, rather than angular, size of the flat part of the column density profile, and therefore any uncertainties in the distance of the cloud will also propagate to the estimate of the central volume density. 
\end{itemize}

Depending on the detailed conditions in the parent coud and the details of the formation mechannism, the structure of the core at radii significantly larger than the central flat part of the column density profile can differ appreciably from the profiles presented here. However, the density profile in the most central parts of the core that we consider here, and at central densities in excess of $\sim 10^5 {\rm \, cm^{-3}}$, has lost, to a large extent, memory of the details of the core formation mechanism and of the physical conditions at the envelope. This has been shown by, e.g., Eng (2002) for the presence of MHD waves in the parent cloud, Tassis \& Mouschovias (2007) for varying amounts of initial magnetic support, and  Gong \& Ostriker (2011) in the case of cores created by colliding turbulent flows. The latter, once collapse starts, exhibit density profiles compatible with models of pure hydrodynamical collapse (Larson-Penston solutions). For this reason, and as long as the analysis is performed very close to the central part of the core, our results are relatively robust to the nature of the processes which are operating at larger scales and which are responsible for the formation of the core. 

Although more detailed modeling can yield more accurate estimates of the volume density and an array of other parameters, the remarkable simplicity of our proposed technique and its insensitivity to the details of the adopted dynamical model render it a fast and useful tool. 

We have verified that the results obtained in well-studied molecular cloud cores L1544 and B68 are consistent with other, more detailed estimates available in the literature. 

\acknowledgements{ We thank the referee Dr. Derek Ward-Thompson for his insightful comments. This work was carried out at the 
Jet Propulsion Laboratory, California Institute of Technology, under a contract with the National Aeronautics 
and Space Administration.  \copyright 2010. All rights reserved.}

\end{document}